\documentclass[]{spie}  

\newcommand{\ket}[1]{|#1\rangle}
 
\usepackage{amsmath,amsfonts,amssymb}
\usepackage{graphicx}
\usepackage[colorlinks=true, allcolors=blue]{hyperref}

\title{Automated Quantum Oracle Synthesis with a Minimal Number of Qubits}

\author[a]{Jessie M. Henderson}
\author[a]{Elena R. Henderson}
\author[a]{Aviraj Sinha}
\author[a]{Mitchell A. Thornton} 
\author[b]{\hspace{3em} D. Michael Miller}
\affil[a]{Southern Methodist University, 6425 Boaz Lane Dallas, TX 75205, USA}
\affil[b]{University of Victoria, P.O. Box 3055 Victoria, BC V8W 3P6, Canada}

\authorinfo{Further author information: (Send correspondence to JMH.)\\JMH: hendersonj@smu.edu, ERH: erhenderson@smu.edu, AS: avirajs@smu.edu,\\ MAT: mitch@smu.edu, DMM: mmiller@uvic.ca}

\begin{document} 
\maketitle

\begin{abstract}
Several prominent quantum computing algorithms---including Grover's search algorithm and Shor's algorithm for finding the prime factorization of an integer---employ subcircuits termed `oracles' that embed a specific instance of a mathematical function into a corresponding bijective function that is then realized as a quantum circuit representation.  Designing oracles, and particularly, designing them to be optimized for a particular use case, can be a non-trivial task.
For example, the challenge of implementing quantum circuits in the current era of NISQ-based quantum computers generally dictates that they should be designed with a minimal number of qubits, as larger qubit counts increase the likelihood that computations will fail due to one or more of the qubits decohering.
However, some quantum circuits require that function domain values be preserved, which can preclude using the minimal number of qubits in the oracle circuit.
Thus, quantum oracles must be designed with a particular application in mind.
In this work, we present two methods for automatic quantum oracle synthesis.
One of these methods uses a minimal number of qubits, while the other preserves the function domain values while also minimizing the overall required number of qubits.
For each method, we describe known quantum circuit use cases, and illustrate implementation using an automated quantum compilation and optimization tool to synthesize oracles for a set of benchmark functions; we can then compare the methods with metrics including required qubit count and quantum circuit complexity.
\end{abstract}


\section{Introduction}
\label{introduction}
A quantum circuit is a representation of an algorithm intended to run on a quantum computer using the gate model paradigm of computation~\cite{Fey:82, Deu:85, Deu:89}. 
This model consists of quantum bits, or qubits, and a sequence of low-level quantum operations, or quantum gates, that act upon those qubits.
Graphically, a quantum circuit is depicted as a set of horizontal lines that represent the state of qubits over some time interval, during which gates, represented by symbols intersecting the horizontal lines, act upon the qubits, causing them to evolve to a new quantum state.  While many quantum circuits are well-known patterns, a significant number also require the specification of an arbitrary mathematical function as a subcircuit for a particular use-case.
This requires the quantum programmer to determine a customized subcircuit---termed a quantum oracle---that realizes the particular function of interest.
Furthermore, before producing such an oracle, the quantum programmer must convert each function of interest---which is not necessarily a bijection nor completely specified---to a corresponding bijective, completely-specified `embedding function' that can be represented by an oracle subcircuit.

Manual oracle design becomes challenging as either the cardinality of the dependent variable set or the number of qubits required to represent the function domain increases.
In addition to the challenge of determining an appropriate embedding function, the act of specifying the embedding function as a series of low-level quantum operators quickly becomes an unwieldy task for human programmers.
Oracle synthesis is made even more challenging given the constraints of contemporary quantum hardware, which is part of the noisy, intermediate-scale quantum (NISQ) era~\cite{Pre:18}.
Today's quantum computers must overcome tremendous engineering challenges to prevent qubits from decohering during circuit execution, and thus, even state-of-the-art gate-based machines that have a mere $O(10^2)$ qubits are susceptible to decoherence.
Therefore, it is desirable to minimize both the number of qubits and the number of required gates when designing quantum circuits, adding yet more considerations to a quantum developer's list.

Not only is quantum oracle design a difficult problem, but it is also an important one, because many quantum algorithms include structures that are---or can be considered---quantum oracles.
For example, consider Grover’s database search program, in which an oracle implements a function that indicates if an element within a collection of search objects satisfies a programmer-selected criterion.
Although the characteristics of Grover's search algorithm are well-understood, the structure of the Grover oracle depends upon the types of objects within the database, as well as the characteristics of the sought object.
Therefore, the oracle used within an implementation of Grover’s search must be designed for each use case.

Given the difficulty and importance of oracle synthesis, designing tools to aid quantum software developers with oracle generation is an active area of research.
For example, recent work developed an automated oracle synthesis framework for formally verifying generated oracle circuits using a custom ``oracle quantum assembly language"~\cite{li2022verified}.
In this article, we focus on circuit synthesis rather than formal verification.
We present a tool for automatically synthesizing oracle circuits from user-specified functions, even when those functions are not reversible nor completely-specified.
We thus provide quantum developers a more flexible, natural way of specifying functions of interest, while also relieving them of the difficult task of manually formulating embedding functions or manually translating those embedding functions into a minimal set of qubits and quantum operators.

The automated oracle synthesis tool comprises two different oracle synthesis methods, one of which produces an oracle with the minimum number of required qubits to support the embedding function, and the other of which produces an oracle with more qubits---but usually fewer gates---in order to preserve the embedding function domain values.
The remainder of this paper proceeds as follows.
Section~\ref{background} provides a brief introduction to quantum circuits and quantum oracles, and includes discussion of well-known quantum algorithms that employ oracles.
Section~\ref{function_embedding} describes methods of determining embedding functions.
This is the portion of the synthesis process that allows developers to automatically obtain complete, reversible circuits from incomplete, irreversible function specifications.
Section~\ref{oracle_synthesis_methods} then describes the two oracle synthesis methods implemented in our quantum compilation and optimization tool, before Section~\ref{experimental_results} illustrates our implementation on both a straightforward conceptual example and on benchmark circuits.
We conclude with a brief discussion of potential use cases for the different synthesis methods, as well as avenues for future work.

\section{Background}\label{background}
\subsection{Quantum Circuits and Oracles}\label{background_quantum_circuits_oracles}
\begin{figure} [ht]
\begin{center}
\begin{tabular}{c}
\includegraphics[width=0.5\textwidth]{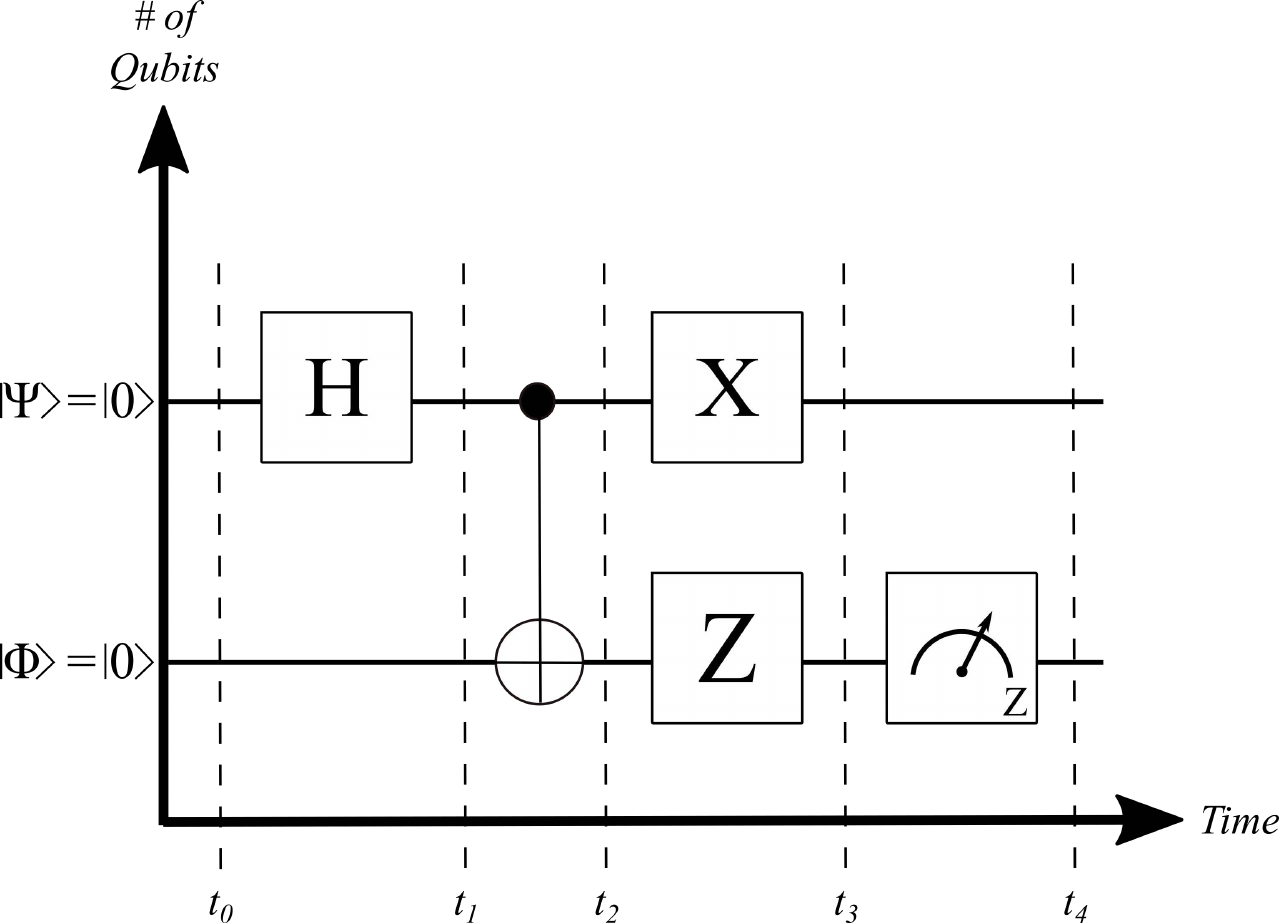}
\end{tabular}
\end{center}
\caption[example] 
{\label{fig:example_circuit} 
An example of a quantum circuit.}
\end{figure}

Figure~\ref{fig:example_circuit} depicts a sample quantum circuit that is annotated with a coordinate system.
This graphical depiction of a quantum program represents the composite quantum state of the qubits at different times, as indicated by the horizontal coordinate axis.
Specifically, Figure~\ref{fig:example_circuit} is comprised of two qubits represented by the two horizontal lines.
The vertical axis depicts the discrete number of qubits in the circuit, and the symbols intersecting the horizontal lines represent quantum state transformations, or gates, applied to the qubits.
For example, the box annotated with \textbf{H} on the leftmost side of the figure represents the application of a Hadamard gate to qubit $\ket{\psi}$ that occurs during the time interval of $(t_0,t_1)$.
And the symbol in the time interval $(t_1,t_2)$ represents a controlled-$\mathbf{X}$ gate, a two-qubit operation applied to the $\ket{\psi\phi}$ qubit pair that is also commonly referred to as a `controlled-NOT' gate.\footnote{This paper uses Dirac notation, which is a standard way of mathematically representing quantum state vectors.  For an introduction to Dirac notation, please see Ref.~\citenum{MSA:22} or Ref.~\citenum{NC:00}.}

Quantum oracles are subcircuits that are used in several well-known quantum algorithms.
When considered as part of such an algorithm, oracles are blackbox functions that act on circuit outputs based upon circuit inputs.
An oracle's action depends both upon the algorithm for which the oracle is being developed (\textit{e.g.}, Grover's search, Simon's periodicity, etcetera) and upon the specific context in which the algorithm's task is being performed (\textit{e.g.}, for Grover's algorithm, searching a deck of cards, searching a database of employees, etcetera).
Thus, quantum algorithms can be developed `around' such oracles, allowing for consideration of the overall process without requiring synthesis of a particular oracle for a particular contextual case.

For example, consider Grover's search algorithm, which uses an oracle to identify database items that satisfy user-specified constraints.
The exact gates constituting this oracle must change to fit each different database and search query.
However, the overall \textit{structure} of the Grover search algorithm involves querying an oracle, for which the precise gates are unspecified, but for which functionality is understood as determining the input values that satisfy a particular search criteria.

For clarity, we provide formal definitions in~\ref{def:reversible_logic} and~\ref{def:switching_func} to clearly differentiate the meaning of the terms `reversible logic' and `switching function' as used in the remainder of this paper. 
\begin{definition}\label{def:reversible_logic}
``Reversible logic" describes a function that is defined over a digit set with a cardinality of two (2). 
Reversible logic functions are restricted to the set of functions that are bijective relations; therefore, a reversible logic function can be specified as a unitary transformation matrix $\mathbf{U}$ that is in the form of a permutation matrix.
Reversible logic functions may be defined over the digit set $\mathbb{Z}_2 = \mathbb{B}=\{0,1\}$ or over the computational basis set $\{\ket{0}, \ket{1}\}$, depending upon the circuit technology for which they are mathematical models.
It is worth noting that irreversible operators such as the switching logic operators OR, AND, XOR, etcetera can be employed in the specification of binary-valued reversible logic functions.
\hfill $\Box$
\end{definition}

\begin{definition}\label{def:switching_func}
A ``switching function" is a representation of any function wherein the range consists of a subset of elements from the set $\mathbb{Z}_2^m$ and, likewise, the domain is a subset of elements from the set $\mathbb{Z}_2^n$.
The set $\mathbb{Z}_2$ is defined as $\mathbb{Z}_2=\mathbb{B}=\{0,1\}$, and $m$ and $n$ are positive integers.
\hfill $\Box$
\end{definition}

Given Definitions~\ref{def:initial_func} through~\ref{def:oracle_func}, the objective of this work is thus to determine an oracle function $\mathbf{U}_f$ that represents an embedding function, $f_e(x)$ that, in turn, covers the function of interest, $f(x)$ and to automatically produce a quantum circuit that realizes the oracle function transfer matrix, $\mathbf{U}_f$, as comprising an allowable set of quantum gates.
This automated process is designed to be convenient for users in a number of ways, including by automatically converting irreversible, incomplete switching functions of interest into reversible, complete specifications prior to synthesis, and by producing oracle circuits that use the minimal number of required qubits.

\begin{definition}\label{def:initial_func}
A ``function of interest," $f(x)$, is one that is not necessarily bijective nor completely specified, and that may be characterized with a domain set whose members use a different number of digits than the members within the  range set.
Functions of interest are often determined by quantum software developers when implementing quantum circuit patterns for a specific use case.
\hfill $\Box$
\end{definition} 

\begin{definition}\label{def:embedding_func}
An ``embedding function," $f_e(x)$, is a bijective function that contains the same mappings or relations as those specified by the function of interest, $f(x)$.
Reversible functions are necessarily bijective and it is a consequence of the postulates of quantum mechanics and the definition of a gate-based quantum computer that all quantum circuits represent reversible functions.
\hfill $\Box$
\end{definition} 

\begin{definition}\label{def:oracle_func}
An ``oracle function" is a representation of an embedding function $f_e(x)$ in the form of a unitary quantum circuit transfer matrix denoted as  $\mathbf{U}_f$.
Thus, the oracle function can be mathematically described as $f_e(x)$, or by the unitary transfer matrix, $\mathbf{U}_f$ that has $f(x)$ embedded within it. 
Quantum oracle circuits use a subset of input qubits to represent specific valuations of the domain values $x$, and they may also use an additional subset of output qubits to represent valuations of $f(x)$.  Some quantum orcale circuits also comprise ancillary and/or garbage qubits that are required to enable the represented embedding function, $f_e(x)$ to be bijective.
Input ancillary values, $\ket{a}$, and output garbage values, $\ket{g}$, are incorporated into the oracle function to accurately represent the embedding function's bijectiveness.
\hfill $\Box$
\end{definition}

It is worth noting that an embedding function and its corresponding oracle function are homomorphic, meaning that they can be mapped to one another with a bijective operation.
The same is \textit{not} true for the original function of interest, which is not generally isomorphic to either the embedding or the oracle function since it is not necessarily bijective nor completely specified.
Once an initial function is made bijective and completely specified via conversion to an embedding function, a quantum oracle can be used to represent the same information as is contained in the initial function.

\subsection{Examples of Quantum Algorithms with Oracles}\label{background_example_algorithms_with_oracles}
\begin{figure} [ht]
\begin{center}
\begin{tabular}{c}
\includegraphics[width=\textwidth]{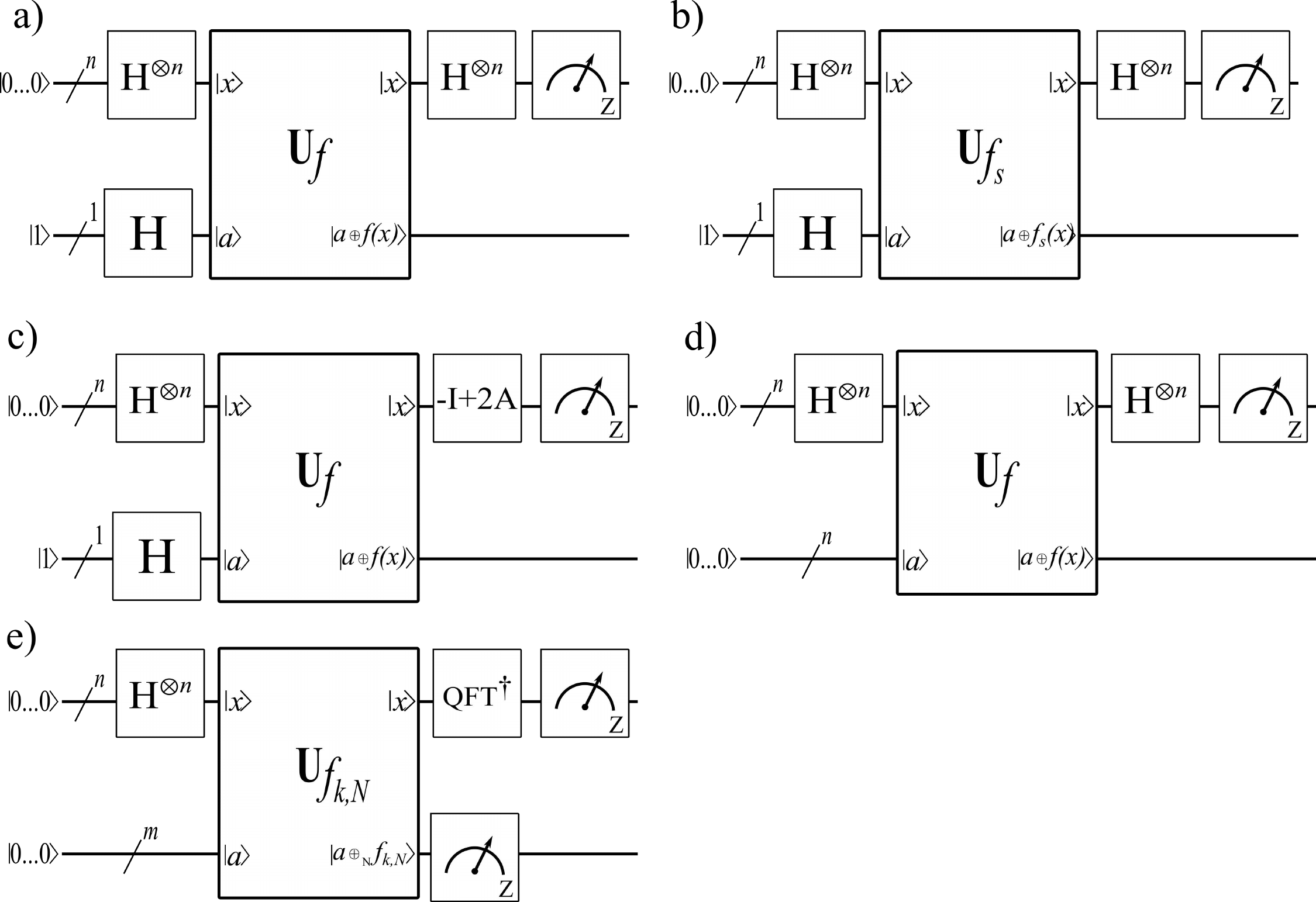}
\end{tabular}
\end{center}
\caption[example] 
{\label{fig:example_oracles} 
Examples of quantum algorithms that use oracles: a) The Deutsch-Jozsa algorithm, b) The Bernstein-Vazirani algorithm, c) Grover's search algorithm, d) Simon's periodicity finding algorithm, and e) Shor's factoring algorithm.}
\end{figure}

To illustrate the purpose---and frequent use---of oracles, we briefly discuss five well-known quantum algorithms that employ oracles.
First is the Deutsch-Jozsa algorithm, which was originally specified as an example to show the power of gate-model quantum computing~\cite{DJ:92, CM+:98}.
The Deustch-Jozsa algorithm embeds an unknown switching function within a ``blackbox" that is implemented as an oracle.
The algorithm's objective is to determine if the embedded function is ``balanced," meaning whether exactly half of the function values evaluate to a unique codomain value.
Subfigure a) of Figure~\ref{fig:example_oracles} depicts the quantum circuit that implements the Deutsch-Jozsa algorithm.

Second is the Bernstein-Vazirani hidden string finding algorithm~\cite{BV:97}.
This algorithm uses an oracle with an embedded function of the form $f_s(x):\mathbb{Z}_2^n \rightarrow \mathbb{Z}_2$, in which some string of length $n$, denoted as $\mathbf{s}$, is referred to as a ``hidden string."
The embedded function $f_s$ is a switching function that yields the inner product of the two $n$-bit strings $\mathbf{x}$ and $\mathbf{s}$, and the Bernstein-Vazirani algorithm computes the hidden string $\mathbf{s}$.
The purpose of this algorithm is to demonstrate properties regarding the complexity classes BQP and BPP.
Subfigure b) of Figure~\ref{fig:example_oracles} contains the quantum circuit implementing the Bernstein-Vazirani algorithm.

Third is Grover’s database search algorithm~\cite{Gro:96}.
As briefly mentioned earlier, the quantum circuit that implements Grover’s search uses an oracle that indicates if a database object is present at the $\ket{x}$ inputs.
Subfigure c) of Figure~\ref{fig:example_oracles} depicts the quantum circuit implementing an iteration of Grover’s search algorithm.

Fourth is Simon’s periodicity finding algorithm~\cite{Sim:97}.
It is similar to the Deutsch-Josza algorithm: the oracle is considered a ``blackbox" that embeds a switching function of the form $f:\mathbb{Z}_2^n \rightarrow \mathbb{Z}_2^n$, and the objective is to determine the period of the function $f$.
Simon’s periodicity finding algorithm is shown in Subfigure d) in Figure~\ref{fig:example_oracles}.

Fifth and finally is Shor’s factoring algorithm that is used to efficiently factor integers into prime factors~\cite{Sho:94}.
The oracle is formed based upon the determination of a modulo-powers function, $f_{k,N}(x)=k^x (\text{mod} N)$, that depends on two constant numeric parameters, $k$ and $N$, which are determined in a preprocessing step.
$f_{k,N}$ serves as the embedded function for the oracle in Shor's algorithm.
Subfigure e) of Figure~\ref{fig:example_oracles} depicts the portion of Shor’s algorithm that executes on a quantum computer.

Before discussing the two forms of quantum oracle that our tool creates, it is worth re-emphasizing that implementing oracles for the above algorithms can require a significant amount of work on the part of quantum software developers when the oracles are manually determined.
In particular, while the function of interest for Shor’s algorithm---the modulo-$N$ powers function---is inherently realizable as a unitary function, the same is not true, in general, for the functions of interest used in the oracles for the Deutsch-Jozsa, Grover’s, and Simon’s algorithms.
Thus---absent a tool such as ours---quantum developers must manually transform such functions to be reversible before synthesising circuits representing them.
Furthermore, regardless of whether the functions are inherently unitary, oracle design always requires a re-synthesis or re-compilation step for each particular situational context, even when the algorithmic use-case remains the same.
Returning to an aforementioned example, performing Grover's searches for two different databases requires two entirely different oracles, as does performing two different searches for the same database of items.
To emphasize the importance of this re-synthesis, we can contrast with classical computing.
In the latter, once an algorithm has been implemented, it requires only a single compilation, even for execution in different situational contexts.
Because the same is not true for quantum algorithms, it is desirable to eliminate the need for a quantum software developer to manually re-synthesize oracle subcircuits for every use case.

\subsection{Quantum Oracles with Minimal Qubits}
\label{background_oracles_minimal_qubits}
Most quantum algorithms use qubits that are defined as quantum states over a binary basis.
For such algorithms, it is necessary to represent a function of interest as a switching function prior to determining the structure of a quantum oracle.
Functions that are not intrinsically defined in the form of a switching function are easily represented in that form by representing all domain and range values as fixed-point signed or unsigned values.
When converting a non-binary function into a switching function form, it is usually the case that a minimal number of bits/qubits, $n$, is used for efficiency.
More specifically, the number of bits or qubits, $n$, used to represent variable valuations, or domain values, $d_i$, in unsigned, fixed-point form is given in Equation~\eqref{eq:num_vars}.

\begin{equation}
\label{eq:num_vars}
n = \lceil \log_2(\max(d_i))\rceil
\end{equation}

Likewise, Equation~\eqref{eq:num_vars_2} is typically used to determine the number of bits or qubits, $m$, representing valuations of a function given a specific variable valuation, or the range values, $r_i$, of a function.

\begin{equation}
\label{eq:num_vars_2}
m = \lceil \log_2(\max(r_i))\rceil
\end{equation}

We note that restricting the switching function form of an embedding function to use these $n$ and $m$ values is a heuristic for reducing the circuit complexity of a resulting quantum oracle circuit.
While using more than the minimally required number of qubits can lead to further reduction of circuit complexity, the determination of an optimal number of required qubits with respect to circuit complexity reduction is an open research question not further addressed in this work.

As previously mentioned, functions of interest need not be bijective, meaning they need not be one-to-one.
Indeed, it is often the case that a specific function has $n\neq m$.
For example, an oracle in a Grover’s search algorithm is usually specified with $m=1$, since the result of the oracle is either ``true" or ``false," depending upon whether or not the search object is present within a larger dataset.
However, quantum program implementations represent unitary transformation matrices, and therefore, quantum oracles need to represent unitary transforms.
As a result, we must generate one-to-one embedding functions from functions of interest that are not intrinsically one-to-one.
Quantum oracles then represent such embedding functions by incorporating ancilla and garbage qubits.
The lower bound for the number of required qubits is given by $N_{min}=max(n,m)+v$, where $v$ is the minimal number of ancilla qubits required by the oracle.
If the function of interest is intrinsically one-to-one, then $v=0$, so  $N_{min}=max(n,m)$.

\begin{figure} [ht]
\begin{center}
\begin{tabular}{c}
\includegraphics[]{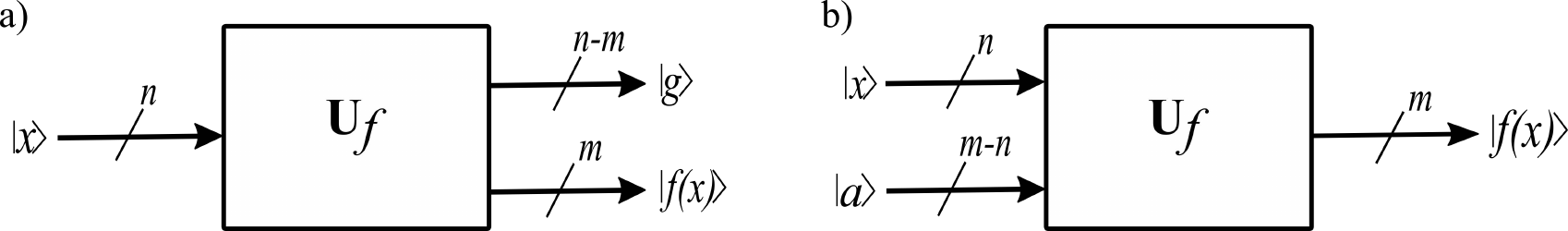}
\end{tabular}
\end{center}
\caption[example] 
{\label{fig:oracle_minimal_qubits} 
An oracle with a minimal number of qubits: $m<n$ (a) and $n<m$ (b).}
\end{figure}

In oracles that minimize qubit count, some of the $n$ qubits that represent the domain values are reused.
In other words, the oracle causes their quantum states to evolve into the quantum states of some or all of the qubits representing the $m$ range qubits.
Specifically, for an embedding function, if $n>m$, then the resulting minimized oracle has $n$ qubits: on the output side, $m$ qubits are used for the range values, and $n-m$ qubits are garbage outputs.
Alternatively, if $n<m$, then the resulting oracle has $m$ qubits: on the input side, $n$ qubits are used for domain values, and $m-n$ ancilla qubits are initialized to some state $\ket{a}$, which is often $\ket{0}$.
Figure \ref{fig:oracle_minimal_qubits} contains a diagram of oracles with these structures.
Each of these implements the oracle function represented by the transfer matrix $\mathbf{U}_f$, which itself represents a unitary transformation containing the embedding function.

\subsection{Quantum Oracles that Preserve the Domain Qubits}
\begin{figure} [ht]
\begin{center}
\begin{tabular}{c}
\includegraphics[]{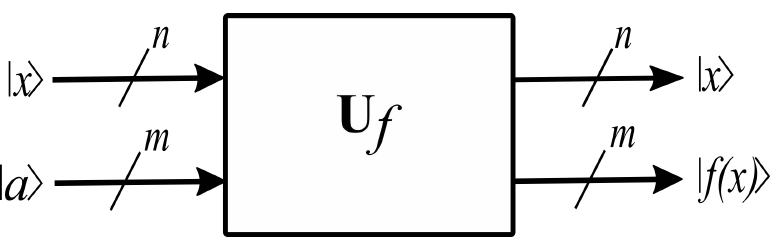}
\end{tabular}
\end{center}
\caption[example] 
{\label{fig:oracle_preserve_domain} 
An oracle that preserves domain values.}
\end{figure}

Many quantum algorithms require the preservation of the domain values after the embedded function is evaluated.
For example, some algorithms---such as Simon's periodicity finding algorithm and Shor's algorithm for determining prime factors---use entanglement properties of the domain values and the corresponding range values (produced by the oracle) for further processing.
Thus, it is important for the oracle to preserve the domain values as part of its output.
Figure~\ref{fig:oracle_preserve_domain} illustrates an oracle with such a domain-preserving structure.

The oracle depicted in Figure~\ref{fig:oracle_preserve_domain} consists of $n+m$ qubits.
The $n$-qubit function variable, $\ket{x}$, and $m$-qubit ancilla, $\ket{a}$, are shown on the input side of the oracle.
The output side consists of the $n$-qubit variable valuation that is identical to the input, $\ket{x}$, and the corresponding function evaluation that is combined in a qubit-wise manner with the $m$ ancilla qubits through the exclusive-OR operator.
Since the oracle, $\mathbf{U}_f$, is a reversible logic function, the exclusive-OR operator operates over the computational basis values of $\ket{0}$ and $\ket{1}$ in the same manner as the conventional digital logic operator defined over elements in $\mathbb{Z}_2$.
It is worth noting that, because the oracle outputs are the $n$ uniquely-valued domain qubits and the $m$ range-valued qubits, the total number of required qubits is always $N_{min}=n+m$.
No additional qubits must be included to serve as ancilla or garbage, even when the function of interest is not one-to-one.
Thus, even when there exist duplicated range values, the presence of the uniquely-valued domain qubits enables the oracle transformation matrix to be unitary.

\section{Function Embedding}\label{function_embedding}
Generating a quantum oracle from a function of interest requires first representing that function in a suitable form.
Specifically, if the function is not given in the form of a switching function, it is first converted to a radix-2 switching function.
For developer convenience, our automated tool represents functions of interest in a tabular form that is straightforward to generate.

We use the \texttt{.pla} file format, which represents switching functions in a table~\cite{BH+:84}.
While a tabular representation permits any function to be specified, such representation is exponential in size, requiring a table with $O(2^n)$ rows and $O(n+m)$ columns of values from the set $\mathbb{Z}_2$.
Each row consists of $n+m$ bits, with the first $n$ bits representing one of the unique bitstrings that corresponds to a unique variable valuation, along with a bitstring of length $m$ representing the corresponding function valuation.
Thus, the \texttt{.pla} format includes features to reduce the size of function-specification tables~\cite{BH+:84}.
First, the $n$-bit function variable values that differ in only a single bit value from other variable values are collapsed into a single row, with the ``-" symbol replacing the appropriate bit to indicate that both ``0" and ``1" values are represented.
Table rows may contain multiple instances of the ``-" character, combining even more rows.
Second, rows are not included for undefined range values.
Third, if certain function range values may be assigned either a ``0" or ``1" value, a ``-" is included in the corresponding location of the $m$-bit string; in this case, the ``-" symbol represents a ``don’t care" condition, meaning that either a ``0" or a ``1" can be assigned.
It is important to note the difference in ``-" meaning when present in the $n$-bit string representing the function variable valuations versus the $m$-bit string representing the function value.
In the first case, the ``-" represents that function values exist for \textit{both} a ``0" and a ``1" in place of the ``-", while in the second case, the ``-" represents that \textit{either} a ``0" or a ``1" is a suitable replacement for the ``-".

Before moving on, it is worth noting that---while the \texttt{.pla} format has advantages for reducing the exponential size in many cases---there are alternative representational forms that could be used and that might, in some cases, allow for better table-size reduction than the \texttt{.pla} format.
For example, two other possible representations are text-based structural netlists of conventional digital logic gates using a formally specified language such as Verilog, and graphical forms such as digital logic circuit diagrams or other forms of decision diagrams~\cite{Bry:86}.
The methods described in this work could be adapted to use such forms of representation for embedding functions to leverage situations in which the functions might be more efficiently represented.
However, this work will consider only \texttt{.pla} representations for two reasons: first, the ease of use of \texttt{.pla} files is consistent with developing an oracle-generation tool that is straightforward to use.
Second, the purpose of this work is to describe the methods for efficiently synthesizing quantum oracles, and not to compare different methods for the representation of switching functions.
So, while we acknowledge the possible utility of other representational forms, we leave further exploration to future work.

\subsection{Embedding Functions with Minimal Qubits}\label{embedding_with_minimal_qubits}
When synthesizing a circuit that will use the fewest possible number of qubits, it is required that the circuit be specified as a one-to-one function.
There are a variety of approaches for this task~\cite{MWD:09}, and in this work, we use a three-step process to embed a switching function of interest into a one-to-one representation using a minimal number of inputs and outputs.
This process is termed the RTT method; it was originally presented in a workshop~\cite{GT:18}, and has been refined for use with methods such as those in this work.

The RTT process applies to a given function representation with $n$-bit strings representing function variable values and $m$-bit strings representing function values.
The first step is to determine $N_{dup}$, the maximum number of times that an $m$-bit string is duplicated in the representation.
An $m$-bit string is duplicated when it represents the function value for multiple of the distinct $n$-bit strings that represent function variable values.
The second step of the RTT method uses this value of $N_{dup}$ to add ancilla inputs and garbage outputs for embedding the original function of interest in a one-to-one representation.
If $N_{dup}=0$, then the function of interest is one-to-one, so there are no ancilla or garbage to be added.
Otherwise, $v$ garbage outputs are required, where the value of $v$ is given by $v = \log_2 (N_{dup})$.
Adding $v$ garbage outputs may require the addition of ancilla such that the function remains one-to-one.
Thus, $w$ is the number of ancilla qubits to be added, and is given by $w = v + m - n$.
When added, the $v$ garbage outputs and the $w$ ancilla values are initialized to the value $0$.

The third and final step of the RTT method is to assign  values from $\mathbb{Z}_2$ to the $w$ ancilla inputs and $v$ garbage outputs such that the duplicated $m$-bit function-value strings of the original representation are now differentiated.
For each group of $m$-bit function-value duplicates, the first duplicate instance's garbage remains assigned to $0$, and each subsequent duplicate instance is assigned to the value of the previous instance incremented by $1$.
The same holds for the $n$-bit function values associated with these function valuations: the first instance's ancilla remains assigned to $0$, with each subsequent instance being assigned to a value of the previous instance incremented by $1$.
At this point, the embedding function is represented as a one-to-one function that, when synthesized, will use the minimally-required number of qubits.

\subsection{Embedding Functions that are Mathematically Onto}
The TBS method that we apply to synthesize oracles with minimal qubits requires that the functions be not only one-to-one, but also onto.
There are many approaches for embedding non-onto functions into onto variants~\cite{MWD:09}.
Here, we discuss two methods, both of which are described in Ref.~\citenum{MWD:09}, and both of which are applied after the functions have been made one-to-one, as discussed above.

The first method is a naive approach: there will be an equal number of inputs and outputs that are not specified explicitly as part of the function, and the naive method simply pairs these up randomly.
So, the first input to not have a specified output is paired with the first output to not have a specified input, and so on, until all inputs and outputs are paired.

While there is nothing incorrect about this approach, it is not optimized for the TBS synthesis method.
The TBS algorithm tends to produce circuits with lower quantum circuit complexity---as described in section~\ref{experimental_results}---when the number of input values that are identical to their associated output values is maximized.
Therefore, a more sophisticated approach for generating onto functions that will be synthesized using TBS is to pair as many unspecified input values to matching output values as possible.
We accomplish this in two steps: first, all unspecified input values are matched with identical unspecified output values, assuming that such an output value is available.
Second, once all possible input values have been matched with identical output values, we assign each remaining unspecified input value to an unspecified output value that is as similar to the input value in terms of matched bits (\textit{i.e.}, in terms of Hamming distance) as possible.
For example, consider a function with $n=m=3$ for which $001$ is an unspecified input value that could be paired with either $110$ or $011$, which are unspecified output values.
In this case, $001$ would be paired with $011$, because those values differ by only one bit, while $001$ and $110$ differ by three bits.

\subsection{Embedding Functions with Preserved Domain Qubits}\label{embedding_with_preserved_domain_qubits}
To preserve the values of the domain qubits, the oracle must evolve the $n$ domain values at the quantum circuit inputs to $n$ qubits at the output.
Thus, the function embedding step replaces the function of interest (with $n$-qubit domain values and $m$-qubit range values) with a function that depends upon the same $n$-qubit domain values, but that has $(m+n)$-qubit range values wherein the $n$ domain qubits are now added to the original $m$ range-value qubits.
Fortunately, explicit copying and modification of the tabular representation of the function is not required, and instead occurs inherently by simply considering the table as representing the embedding function as described above: the domain values are the first $n$ qubits of each row in the table plus $m$ zero values, and the range values are the $(n+m)$ values in each row.

\section{Oracle Synthesis Methods}\label{oracle_synthesis_methods}
Before describing the oracle synthesis methods in more detail, we note that---for both methods---the synthesized oracle function is in the form of a technology-independent QASM specification~\cite{CB+:17,CB+:20}.

\subsection{Minimal Qubit Quantum Oracle Synthesis}\label{minimal_qubit_quantum_oracle_synthesis}
Given a completely specified reversible function, Transformation-Based Synthesis (TBS) methods produce a reversible circuit with a minimal number of qubits that realizes the given function.  
The circuit produced is a reversible logic gate cascade comprised of Pauli-$\mathbf{X}$ gates (\textit{i.e.,} NOT), Feynman gates (\textit{i.e.,} controlled-$X$ or CNOT), Toffoli gates (\textit{i.e.,} controlled-controlled-$\mathbf{X}$ or C-CNOT), and multiple-control Toffoli (MCT) gates (\textit{i.e.,} gates with three or more control inputs and a target Pauli-$\mathbf{X}$ operation)~\cite{MMD:03}.

The basic unidirectional TBS method processes a tabular specification by adding gates to the output side of the circuit so that each output pattern matches the corresponding input pattern~\cite{MMD:03}.  
Since the specification is reversible, one could consider the inverse specification, derive a reverse circuit, and then choose whichever produced circuit is the smaller. 
A more effective approach is to apply the method in both directions simultaneously, the so-called bidirectional TBS, choosing to add gates at the input side or the output side of the circuit at each step of the synthesis process~\cite{MMD:03}.
A further extension of TBS to a so-called multidirectional approach is described in Ref.~\citenum{SDRM:16}, and search based TBS methods have been described in Ref.~\citenum{MD:20}. 
Complete details of both basic and varied TBS methods are provided in the above references; here, we provide a summary of the basic TBS method.

The input to basic TBS is a truth-table specification of a completely-specified, reversible (bijective) switching function with the number of inputs ($n$) equal to the number of outputs ($m$).  
For example, a tabular specification of a function produced using the RTT method described in Section~\ref{embedding_with_minimal_qubits} is a valid input for TBS.
Given such a specification, the basic TBS method identifies the desired reversible circuit by iteratively transforming the given reversible function into the identity function and obtaining a cascade of gates that implements the required transformation.
Reversing that cascade identifies a circuit that maps the identity function to the specified reversible function, and thus maps each input pattern to the required output pattern.  
The gate cascade identifies a quantum circuit, and for our purposes, a quantum oracle.

The basic TBS method works through the given tabular specification in ascending order of the input patterns, \textit{i.e.,} $(0...00), (0...01), (0...10) ... (1...11)$.
Note that for ease of implementation, a given tabular representation is typically reordered so that the input patterns are in the desired order.
At each step, gates are chosen to map the given output pattern to one that matches the input pattern, and are added to the output side of the cascade being constructed.
The gates chosen are applied to the function being synthesized, resulting in an intermediate residual function, $f_{int}$, that becomes the function to be synthesized in the next step.
In other words, each time a reversible gate is determined and appended to the output specification, a new intermediate representation of the function to be synthesized, $f_{int}$, is created for use in subsequent TBS operations.

Consider the first step of basic TBS, in which the input pattern is $(0...0)$, and a set of NOT gates is chosen to make the output pattern $(0...0)$.
In each subsequent step, gates are chosen to map the given output pattern to the target input pattern---the key being that those gates are chosen so that they do not affect any entry earlier in the specification~\cite{MMD:03}.
After the second-to-last row of the specification has been transformed, the last row must be such that the input and output patterns are both $(1...11)$ and no further gates are required. 
The $f_{int}$ function will at that point be the identity function, and, as noted earlier, the reverse of the gate sequence that produced the identity represents the synthesized gate cascade for the initial given reversible function.

Finally, we note that while the initial specification of basic TBS in Ref.~\citenum{MMD:03} and the description above are stated in terms of a truth-table specification, the method can also be implemented using more compact representations such as decision diagrams.

\subsection{Oracles that Preserve the Variable Values}
When an oracle should preserve the variable values by producing them as oracle outputs, the RTT procedure need not be applied to the function of interest prior to synthesis.
As described in Section~\ref{embedding_with_preserved_domain_qubits}, the function of interest is implicitly converted to an embedding function by interpreting each domain and range valuation as a combined $n+m$ output, thus resulting in an oracle that always consists of $n+m$ qubits, with $n$ qubits producing the variable values, and $m$ qubits evolving to a state representing the function values.
Specifically, the \texttt{.pla} table representing the function of interest is used directly as input to a procedure termed the ESOP synthesis method, which was first described in Ref.~\citenum{FTR:07}.
The ESOP method produces an oracle with the structure illustrated in Figure~\ref{fig:oracle_preserve_domain}.

A function in ESOP form is represented as an exclusive-OR sum-of-products: function variables are combined into products via logical conjunctive AND operators, and these products are summed together with disjunctive exclusive-OR operators~\cite{FTR:07}.
When the resulting expression is evaluated at variable values, it produces the corresponding function values.
A \texttt{.pla} table can be converted into ESOP form, meaning it comprises a list of products that produce the ESOP representation of the function when combined via exclusive-OR operators.
Each row of input variable values of the table represents a product, wherein a value of $1$ or $0$ corresponds to an input in positive or negative polarity, respectively.
In our synthesis tool, we apply an implementation of EXORCISM-4 to obtain \texttt{.pla} tables in a minimized ESOP form~\cite{mishchenko2001fast}.

For each product in a given ESOP-form \texttt{.pla} table, the ESOP method produces a Toffoli gate, and thereby maps a list of products to a reversible logic circuit~\cite{FTR:07}.
The basic steps for this mapping are as follows.
For each input variable and each output variable of the ESOP-form \texttt{.pla} table, a qubit is added to the generated circuit as an input or output, respectively.
For each product of the table, each output variable value of $1$ maps to a Toffoli gate with its target on the corresponding output qubit.
The controls of each Toffoli gate are determined by the inputs corresponding to the output variable.
Specifically, an input variable value of $1$ means a control is placed on the corresponding input qubit, while an input variable value of $0$ means a control is placed on the corresponding input qubit along with two Pauli-$\mathbf{X}$ gates, one on either side of the control, in order to reverse the qubit's polarity and then restore it for subsequent Toffoli gates.
This process produces a circuit of $n+m$ qubits, and thus, given a \texttt{.pla} table representing a function of interest interpreted as an embedding function, the ESOP method produces an oracle that preserves variable values.

\section{Experimental Results}\label{experimental_results}
We implemented the two oracle synthesis methods described in Section~\ref{oracle_synthesis_methods} in a quantum synthesis, compilation, and optimization tool termed \textit{MustangQ}~\cite{ST:17, ST:19a, ST:19b, SHT:22}.
\textit{MustangQ} is designed to use various internal representations of both conventional switching functions and quantum circuits, including \texttt{.pla}, QMDD~\cite{MT:06}, and a customized data structure that represents the structural form of a quantum circuit comprised of gates.

To illustrate \textit{MustangQ}'s automated synthesis of oracle functions, we first consider a small and conceptually-straightforward example: synthesizing a Grover's search oracle for a database containing a `deck of cards.'
We then extend this example by illustrating how the output of \textit{MustangQ} can be used to run simulations of Grover's algorithm using Qiskit.
Thus, this example not only illustrates the oracle-generation aspect of \textit{MustangQ}, but also the tool's utility when preparing circuits for specific quantum hardware.
After the deck of cards example, we illustrate \textit{MustangQ}'s data generation using both the ESOP and TBS methods on a series of benchmark functions specified in \texttt{.pla} format.
This allows us to compare the two methods, highlighting the benefits of each.

\subsection{Starting Simple: Oracles for Searching a Deck of Cards}
Consider a database comprising a representation of the fifty-two cards in a standard playing deck.
Each card is specified as a six-bit switching function, where the first two qubits encode the suit, and the latter four qubits encode the numerical value of the card.
Table~\ref{table:cardendcoding} illustrates the card encoding; note that three of the numerical values are unused, because they do not correspond to the numerical values 2-10, Ace, Jack, Queen, or King.

 \begin{table}[ht]
    \label{table:cardendcoding}
     \caption{Encoding card suit as a two-digit bitstring and card rank as a four-digit bitstring.}
    \hspace*{4cm}
    \begin{tabular}[t]{|c|c|}
        \hline
         Clubs & 00\\ \hline
         Spades & 01 \\ \hline
         Diamonds & 10\\ \hline
         Hearts & 11 \\ \hline
         Ace & 0001\\ \hline
        2 & 0010 \\ \hline
        3 & 0011 \\ \hline
        4 & 0100 \\ \hline   
        5 & 0101 \\ \hline
    \end{tabular}
    \hfill
    \begin{tabular}[t]{|c|c|}
        \hline
        6 & 0110 \\ \hline
        7 & 0111 \\ \hline
        8 & 1000 \\ \hline
        9 & 1001 \\ \hline
        10 & 1010 \\ \hline
        Jack & 1011 \\ \hline
        Queen & 1100 \\ \hline
        King & 1101 \\ \hline
        Unused & 0000, 1110, 1111 \\ \hline
    \end{tabular}
    \hspace*{4cm}
\end{table}

Before considering a specific example with the playing cards, we briefly consider Grover’s algorithm in general.
As shown in Figure~\ref{fig:example_oracles}, there are three segments of a circuit implementing the search.
First, all qubits are put into superposition to ensure that all possible items in the search space are considered.
Second, the oracle `marks' one or more desired items by changing the phase of their associated states to be opposite that of the other states in the search space.
Third, `inversion about the mean' magnifies the amplitudes of the `marked' states, while reducing the amplitudes of unmarked states, and then inverts all of the amplitudes about the average amplitude.
Repeated application of the oracle and the inversion about the mean results in a circuit that---when its qubits are measured---is likely to collapse to any of the desired states.
The number of repetitions required is proportional to the \textit{square root} of the number of elements in the search space, and this provides the speedup over classical methods, which have a complexity on the order of the number of elements in the database.

The portions of the circuit for both the first and third steps are entirely determined by the number of qubits present, which is in turn determined by the size of the database.
Conversely, the second portion specifying the oracle changes based upon both the database being searched and the particular search query, so it must be re-synthesized for each new search.
\textit{MustangQ} automates this synthesis for different searches, as is illustrated by the following oracles synthesized to find a specific card or subset of cards in the 64-item deck-of-cards database.

\begin{figure} [ht]
\begin{center}
\begin{tabular}{c}
\includegraphics[]{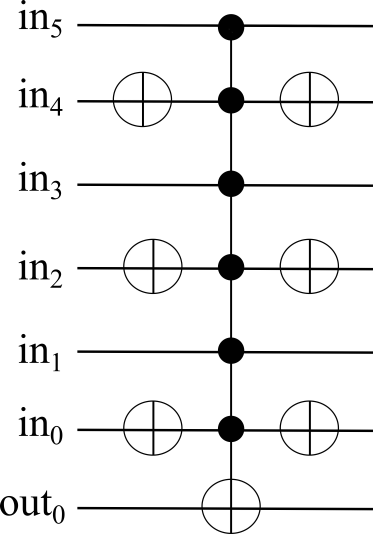}
\end{tabular}
\end{center}
\caption[example] 
{\label{fig:grover_card_example_oracle} 
Oracle for marking the ten of diamonds, encoded as 101010 in our deck-of-cards database.}
\end{figure}

Consider searching for the ten of diamonds, which is equivalent to searching for the state 101010.
Figure~\ref{fig:grover_card_example_oracle} illustrates the oracle that \textit{MustangQ} generated to mark the state 101010 as the sought member of the database.
In this straightforward situation, the oracle is trivially simple, because it involves solely a generalized-Toffoli gate with controls on each of the six input qubits, and a target on the output qubit.
The controls are positive for inputs that are designated as one (1) in this case, and are negative for inputs that are designated as zero (0); as illustrated in the figure, `negative' controls are equivalent to positive controls with a Pauli-$\mathbf{X}$ gate on either side.

\begin{figure} [ht]
\begin{center}
\begin{tabular}{c}
\includegraphics[width=\textwidth]{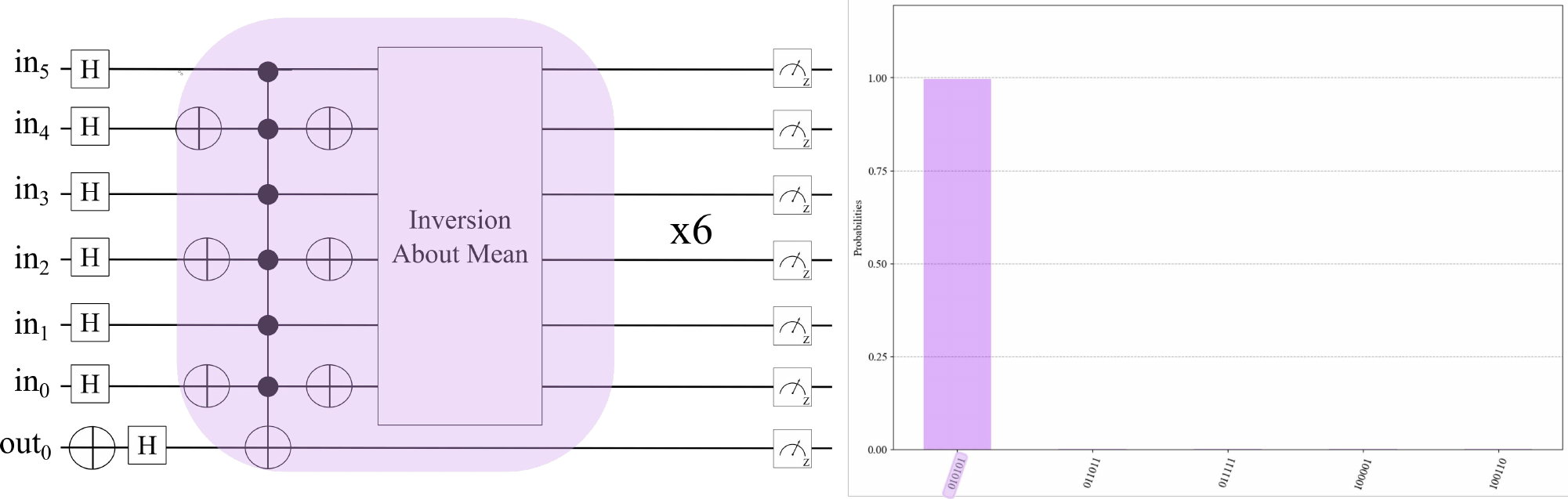}
\end{tabular}
\end{center}
\caption[example] 
{\label{fig:grover_card_example_diamonds} 
The circuit and simulation results for obtaining the ten of diamonds (101010) using Grover's algorithm on an ideal simulator with 1024 individual circuit executions. The results illustrate that the correct result was found in 99\% of the 1024 individual circuit executions.  Note that Qiskit lists qubit results in an order \textit{opposite} that expected, meaning that 101010 appears as 010101 on the $x$-axis.}
\end{figure}

Inserting this oracle into a circuit for the remainder of Grover’s algorithm generates the circuit depicted in Figure~\ref{fig:grover_card_example_diamonds}.
We note two features of this circuit.
First, we re-emphasize that the subroutine denoted ``Inversion About Mean" depends \textit{solely} on the number of qubits present and thus does not vary by database as does the oracle.
Second, we note that the highlighted portion is repeated six times prior to measurement; we determined the number of iterations experimentally, as this trivial example is too small to meaningfully use the recommendation of $O(\sqrt{\frac{N}{M}})$ iterations, where $N$ is the number of items in the database, and $M$ is the number of database items that are solutions to the search query.

After generating the circuit for searching the deck for a ten of diamonds, we used Qiskit to simulate running the circuit on quantum hardware.
Using an ideal simulator (\textit{i.e.,} one that does not introduce noise as a result of imperfect hardware), we generated the results in Figure~\ref{fig:grover_card_example_diamonds}, which shows correctly obtaining the ten of diamonds state (101010) in 99\% of 1024 individual circuit executions.\footnote{IBM terms ``individual circuit executions'' as ``shots'' in their documentation.}

\begin{figure} [ht]
\begin{center}
\begin{tabular}{c}
\includegraphics[width=1\textwidth]{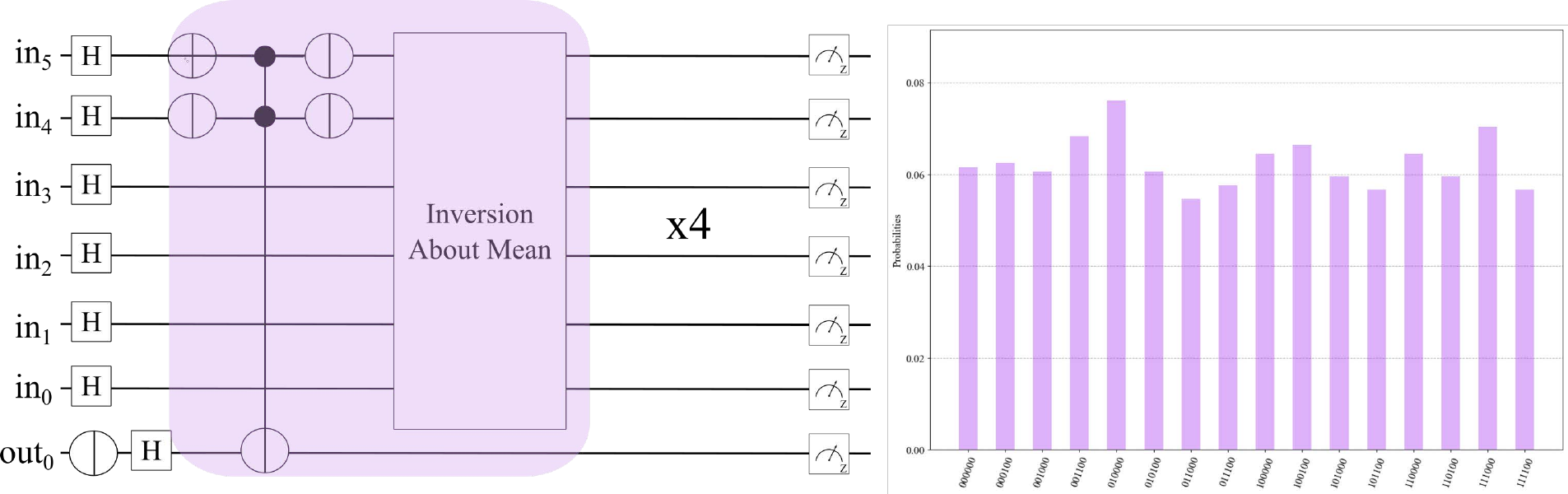}
\end{tabular}
\end{center}
\caption[example] 
{\label{fig:clubs_circuit_and_simulation} 
The circuit and simulation results for obtaining all cards of the club suit (with first digits 00) using Grover's algorithm on an ideal simulator with 1024 individual circuit executions. Note that Qiskit lists qubit results in an order \textit{opposite} that expected, meaning that all results with the clubs suit have 00 as their last two qubit values.}
\end{figure}

Figure~\ref{fig:clubs_circuit_and_simulation} illustrates another search for any card with a clubs (00) suit.
While the concepts of the oracle design and the oracle's inclusion in the Grover circuit are the same as for the ten of diamonds case, there is an important difference: instead of seeking only one card, we expect to find every card (\textit{i.e.}, bitstring) with a 00 as the suit.
Figure~\ref{fig:clubs_circuit_and_simulation} illustrates that our results are as expected: we measure sixteen different bitstring encodings, each of which is either the Ace through King of clubs or one of the three `unused' numerical values (0000, 1110, 1111) for the clubs suit.

\subsection{Oracles for benchmark functions}
The oracles for the deck-of-cards searches provide a straightforward conceptual example of oracle synthesis.
In this section, we illustrate \textit{MustangQ}'s oracle synthesis of complex functions that do not lend themselves to such `manual' consideration.
Due to the limitations of contemporary quantum computers, the circuits we present here are too large to be run on existing machines.
Thus, these examples demonstrate \textit{MustangQ}’s automated oracle synthesis for future quantum hardware and---in the meantime---allow for comparing the two synthesis methods at scale.

\begin{figure} [hbp]
\begin{center}
\begin{tabular}{c}
\includegraphics[width=0.4\textwidth]{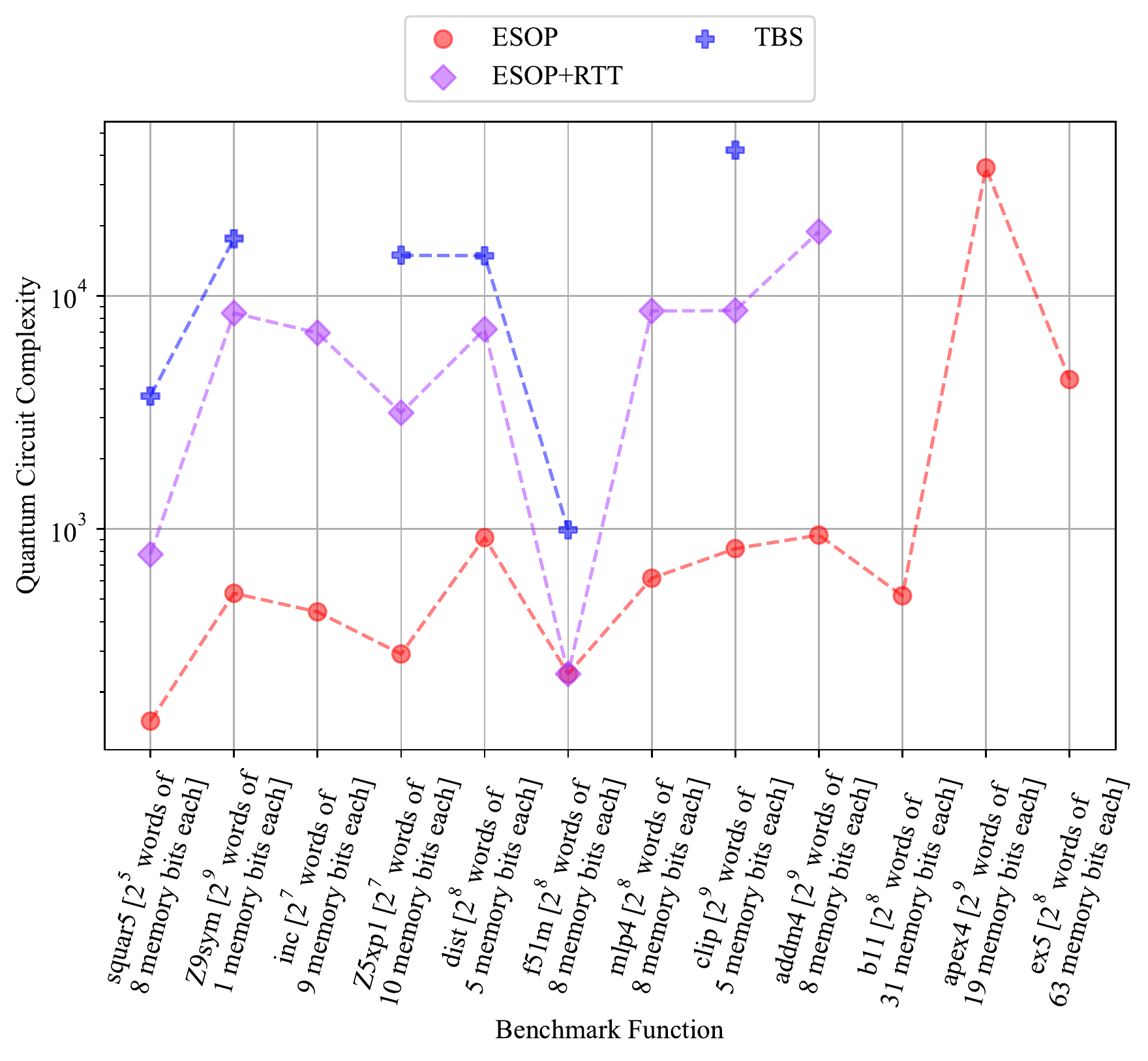}
\end{tabular}
\end{center}
\caption
{\label{fig:quantum_cost_benchmark_results} 
The circuit complexity of twelve \texttt{.pla} functions when synthesized using ESOP synthesis (red circles), ESOP synthesis with non-required RTT processing (purple diamonds), and TBS synthesis (blue crosses).  Some functions were too large to be expanded during RTT or to be addressed with TBS, which has an exponential algorithmic complexity; for these functions, \textit{MustangQ} timed out before completion, so no data point is shown.}
\end{figure}

\begin{figure} [htbp]
\begin{center}
\begin{tabular}{c}
\includegraphics[width=0.4\textwidth]{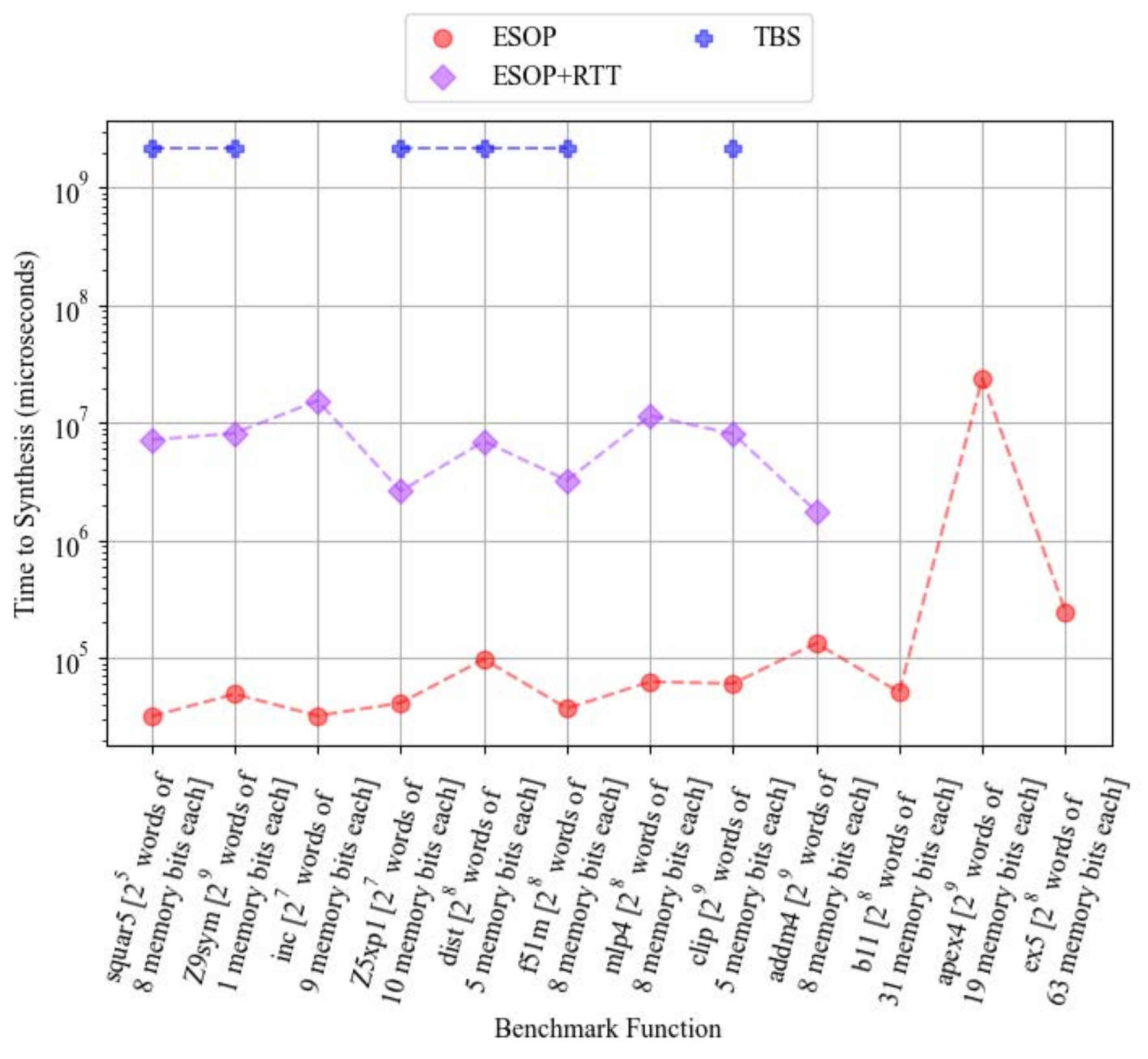}
\end{tabular}
\end{center}
\caption
{\label{fig:time_to_synthesis_benchmark_results} 
The time-to-synthesis (in microseconds) of twelve \texttt{.pla} functions when synthesized using ESOP synthesis (red circles), ESOP synthesis with non-required RTT processing (purple diamonds), and TBS synthesis (blue crosses).  Some functions were too large to be expanded during RTT or to be addressed with TBS, which has an exponential complexity; for these functions, \textit{MustangQ} timed out before completion, so no data point is shown.}
\end{figure}

\begin{figure} [htb]
\begin{center}
\begin{tabular}{c}
\includegraphics[width=0.4\textwidth]{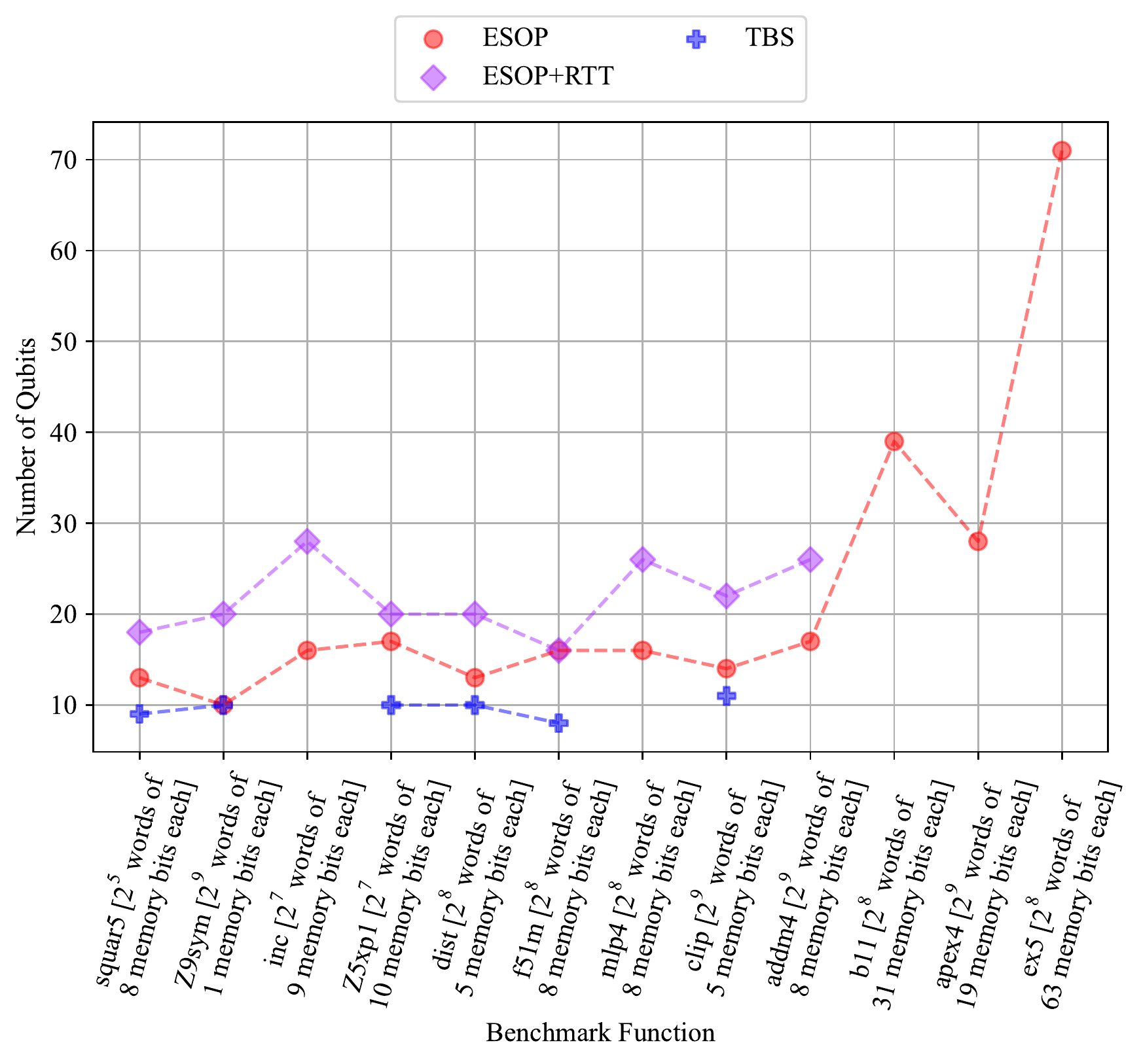}
\end{tabular}
\end{center}
\caption
{\label{fig:qubit_benchmark_results} 
The qubit count of twelve \texttt{.pla} functions when synthesized using ESOP synthesis (red circles), ESOP synthesis with non-required RTT processing (purple diamonds), and TBS synthesis (blue crosses).  Some functions were too large to be expanded during RTT or to be addressed with TBS, which has an exponential complexity; for these functions, \textit{MustangQ} timed out before completion, so no data point is shown.}
\end{figure}

For oracle specifications, we adapt functions from the \texttt{.pla} benchmark set and synthesize three sets of oracles: (1) with ESOP synthesis; (2) with ESOP synthesis following RTT preprocessing; and (3) with TBS synthesis, which requires RTT preprocessing.
(``RTT preprocessing" refers to applying the RTT method to the \texttt{.pla} table prior to synthesis.)
It is worth emphasizing that the combination of ESOP synthesis with RTT preprocessing is for experimental purposes only, meaning it is included not to introduce a practical method of synthesis, but to illustrate a more equitable comparison of the ESOP and TBS methods.
The functions used for each of these three sets are listed in Tables~\ref{table:esop_no_expansion_results},~\ref{table:esop_expansion_results}, and~\ref{table:tbs_results} (in Sec.~\ref{appendix}), and Figures~\ref{fig:quantum_cost_benchmark_results} through~\ref{fig:qubit_benchmark_results} illustrate the results.

For each quantum circuit, we collected the time-to-synthesis, as well as three metrics for quantum circuit evaluation: qubit count, gate count, and a circuit complexity metric that is representative of quantum cost~\cite{smith2019quantum}.
Qubit and gate count are self-explanatory.
By contrast, quantum cost is not standardized; it is proportional to the number of gates in a circuit, but weights gates differently according to their properties~\cite{maslov2005comparison}.
Consequently, the quantum cost of a circuit is technology-dependent, depending upon the native gate set of the machine on which a circuit is run.
However, regardless of the precise weighting of the gates, more complex circuits---meaning those including more gates that act on more qubits---will have larger cost values than circuits that are simpler---meaning with fewer gates that generally act on fewer qubits.
Therefore, we assess quantum circuit complexity using a straightforward, technology-independent metric: the circuit complexity is the sum of the `costs' of each gate, where gate `cost' is the number of qubits on which a gate acts, including both control and target qubits.

In the following discussion, we compare the synthesis methods across three metrics: circuit complexity (which exhibits the same trends as quantum gate count), time-to-synthesis, and qubit count.
First, consider circuit complexity.
As illustrated in Figure~\ref{fig:quantum_cost_benchmark_results}, the functions synthesized with the ESOP method (with no RTT preprocessing), generate oracles with an average circuit complexity of 3,769.5.
By contrast, TBS synthesis produces an average circuit complexity of 15,698.5.
This increase in circuit complexity is likely explained by ESOP's ability to take advantage of incompletely-specified functions, which TBS currently cannot.
As implemented in EXORCISM-4, the process of \texttt{.pla} table conversion to ESOP form can take advantage of ``-" symbols to produce a minimal ESOP representation, which reduces the number of gates---and thus the circuit complexity---produced by ESOP synthesis~\cite{mishchenko2001fast}.
TBS synthesis does not have this advantage, since it requires fully-specified \texttt{.pla} representations, which we obtain via the RTT method.
Specifically, RTT preprocessing results in fully expanded and assigned \texttt{.pla} representations, meaning ``-" symbols present in the inputs are removed, and every output---including those where all functional values are $0$---is represented explicitly.
Figure~\ref{fig:quantum_cost_benchmark_results} further supports this hypothesis by illustrating that the circuit complexity is significantly higher (on average, 6,981.6) when ESOP is applied after RTT preprocessing than when ESOP is applied without RTT preprocessing.
This again suggests that function expansion is one cause of TBS' higher circuit complexities. 

Fully-explicit function representations not only result in increased circuit complexity, but also in increased time-to-synthesis, which is again higher for TBS than for ESOP.
(See Figure~\ref{fig:time_to_synthesis_benchmark_results}.)
On average, synthesis time was $2.15e9$ microseconds for TBS, while it was $7.72e4$ microseconds for ESOP.
(Note that the average for ESOP does not consider the outlier of \texttt{apex4}, which was too large for synthesis with TBS.)
Additionally, as with circuit complexity, ESOP synthesis with RTT preprocessing exhibits higher synthesis times than for ESOP alone; the average time-to-synthesis becomes $7.23e6$ microseconds, which is two orders of magnitude higher than ESOP's average, and three orders of magnitude lower than TBS'.

The trend is different, however, when comparing qubit count: TBS synthesis is the clearly preferred method here, regardless of whether RTT preprocessing is used with ESOP synthesis or not.
(See Figure~\ref{fig:qubit_benchmark_results}.)
Rounding upward to the nearest qubit, TBS synthesis produces oracle circuits with an average of 10 qubits, while those of ESOP synthesis have averages of 22 or 23 qubits, with and without RTT preprocessing, respectively.
This is a reduction of over $56\%$ in terms of qubit count.
Thus, we see that---in their current forms---the algorithms provide different advantages: TBS synthesis improves qubit count, while ESOP synthesis improves circuit complexity and synthesis time.
As we hypothesize the current benefits of ESOP lie in its ability to leverage incompletely-specified functions, we are exploring modifications of TBS that would preserve the minimal qubits while also reducing circuit complexity and synthesis time.
Specifically, we are working on modifying TBS synthesis such that it can work with incompletely-specified \texttt{.pla} representations, thus avoiding the table expansion and assignment that seems to increase circuit complexity and certainly increases synthesis time.

\section{Conclusion and Future Work}
This work presents and illustrates two methods for the automatic synthesis of quantum oracles.
One method is based upon minimizing the total number of required qubits, and the other preserves the function domain.
We implemented both methods within the \textit{MustangQ} quantum synthesis and compiler tool, and we provide experimental results illustrating the strengths of both methods.
There are at least three ways that we could further develop this work.

First, as mentioned earlier in the paper, there are a number of ways that functions could be preprocessed to satisfy the synthesis method constraints.
Several methods are described in Ref.~\citenum{MWD:09}, and we could explore how applying different methods affects the circuit complexity of resulting synthesized circuits.
Second, the TBS synthesis method can be enhanced to make use of more compact function representations that might improve the circuit complexity, the speed of processing, or both.
We are actively developing such improvements to be added and tested in \textit{MustangQ}.
Third and finally, the techniques described in this work use a particular form of quantum data encoding referred to as ``basis encoding," which uses a qubit to represent each conventional bit of information~\cite{SHT:22}.
It is known that alternative methods of data encoding can be employed in quantum circuits; for example, a non-binary value can be encoded into a probability amplitude or a phase angle.
As we have shown in other work, such alternative forms of data encoding can result in quantum circuits that comprise fewer qubits or lower circuit complexity than those utilizing basis encoding~\cite{SHT:22}.
Therefore, future work ought to investigate the use of alternative forms of data encoding within oracle synthesis algorithms to explore the possibility of further qubit and circuit complexity reduction.

\pagebreak

\bibliography{report} 
\bibliographystyle{spiebib} 

\section{Appendix: Tables of Benchmark Data}\label{appendix}
\begin{table}[htbp]
\caption{Results for a set of benchmark functions with ESOP synthesis (and with no RTT method preprocessing).} 
\label{table:esop_no_expansion_results}
\begin{center}       
\begin{tabular}{|c|c|c|c|c|c|c|c|}
\hline
\rule[-1ex]{0pt}{3.5ex}  Function & Inputs & Outputs & Qubits & Gate Count & Circuit Complexity & Time-to-Synthesis ($\mu$s) \\
\hline
\rule[-1ex]{0pt}{3.5ex}  squar5	& 5 & 8 & 13 & 52 & 150 & 3.20e4 \\
\hline
\rule[-1ex]{0pt}{3.5ex}  Z9sym & 9 & 1 & 10 & 157 & 530 & 4.95e4  \\
\hline
\rule[-1ex]{0pt}{3.5ex} inc & 7 & 9 & 16 & 118 & 441 & 3.24e4 \\
\hline
\rule[-1ex]{0pt}{3.5ex}  Z5xp1 & 7 & 10 & 17 & 100 & 291 & 4.16e4 \\
\hline
\rule[-1ex]{0pt}{3.5ex}  dist & 8 & 5 & 13 & 220 & 918 & 9.80e4 \\
\hline
\rule[-1ex]{0pt}{3.5ex}  f51m & 8 & 8 & 16 & 88 & 239 & 3.74e4 \\
\hline
\rule[-1ex]{0pt}{3.5ex}  mlp4 & 8 & 8 & 16 & 147 & 615 & 6.30e4 \\
\hline
\rule[-1ex]{0pt}{3.5ex}  clip & 9 & 5 & 14 & 188 & 824 & 6.09e4 \\
\hline
\rule[-1ex]{0pt}{3.5ex}  addm4 & 9 & 8 & 17 & 219 & 942 & 1.34e5 \\
\hline
\rule[-1ex]{0pt}{3.5ex}  b11 & 8 & 31 & 39 & 132 & 517 & 5.14e4 \\
\hline
\rule[-1ex]{0pt}{3.5ex}  apex4 & 9 & 19 & 28 & 5565 & 35393 & 2.38e7 \\
\hline
\rule[-1ex]{0pt}{3.5ex}  ex5 & 8 & 63 & 71 & 756 & 4374 & 2.49e5 \\
\hline
\end{tabular}
\end{center}
\end{table}

\begin{table}[htbp]
\caption{Results for a set of benchmark functions with ESOP synthesis and RTT method preprocessing. Asterisks (*) denote a function that failed to synthesize due to too many inputs for explicit \texttt{.pla} expansion.} 
\label{table:esop_expansion_results}
\begin{center}       
\begin{tabular}{|c|c|c|c|c|c|c|c|}
\hline
\rule[-1ex]{0pt}{3.5ex}  Function & Inputs & Outputs & Qubits & Gate Count & Circuit Complexity & Time-to-Synthesis ($\mu$s) \\
\hline
\rule[-1ex]{0pt}{3.5ex}  squar5	& 5 & 8 & 18 & 176 & 778 & 7.18e6 \\
\hline
\rule[-1ex]{0pt}{3.5ex}  Z9sym & 9 & 1 & 20 & 1255 & 8431 & 8.18e6 \\
\hline
\rule[-1ex]{0pt}{3.5ex} inc & 7 & 9 & 28 & 773 & 6936 & 1.55e7 \\
\hline
\rule[-1ex]{0pt}{3.5ex}  Z5xp1 & 7 & 10 & 20 & 587 & 3150 & 2.63e6  \\
\hline
\rule[-1ex]{0pt}{3.5ex}  dist & 8 & 5 & 20 & 1189 & 7186 & 6.99e6 \\
\hline
\rule[-1ex]{0pt}{3.5ex}  f51m & 8 & 8 & 16 & 88 & 239 & 3.25e6 \\
\hline
\rule[-1ex]{0pt}{3.5ex}  mlp4 & 8 & 8 & 26 & 1092 & 8613 & 1.15e7 \\
\hline
\rule[-1ex]{0pt}{3.5ex}  clip & 9 & 5 & 22 & 1306 & 8645 & 8.08e6 \\
\hline
\rule[-1ex]{0pt}{3.5ex}  addm4 & 9 & 8 & 26 & 2480 & 18856 & 1.77e6 \\
\hline
\rule[-1ex]{0pt}{3.5ex}  b11 & 8 & 31 & * & * & * & * \\
\hline
\rule[-1ex]{0pt}{3.5ex} apex4 & 9 & 19 & * & * & * & *  \\
\hline
\rule[-1ex]{0pt}{3.5ex}  ex5 & 8 & 63 & * & * & * & * \\
\hline
\end{tabular}
\end{center}
\end{table}

\begin{table}[htbp]
\caption{Results for a set of benchmark functions with TBS synthesis, which requires fully-specified bijective \texttt{.pla} representations that we obtain via RTT method preprocessing. Asterisks (*) denote a function that failed to synthesize due to either too many inputs for explicit \texttt{.pla} expansion or too many required qubits/gates for TBS synthesis (more than 50,000 gates).}
\label{table:tbs_results}
\begin{center}       
\begin{tabular}{|c|c|c|c|c|c|c|c|}
\hline
\rule[-1ex]{0pt}{3.5ex}  Function & Inputs & Outputs & Qubits & Gate Count & circuit complexity & Time-to-Synthesis ($\mu$s) \\
\hline
\rule[-1ex]{0pt}{3.5ex}  squar5	& 5 & 8 & 9 & 1463 & 3714 & 2.15e9 \\
\hline
\rule[-1ex]{0pt}{3.5ex}  Z9sym & 9 & 1 & 10 & 6028 & 17589 & 2.15e9 \\
\hline
\rule[-1ex]{0pt}{3.5ex} inc & 7 & 9 & * & * & * & * \\
\hline
\rule[-1ex]{0pt}{3.5ex}  Z5xp1 & 7 & 10 & 10 & 5983 & 14924 & 2.15e9 \\
\hline
\rule[-1ex]{0pt}{3.5ex}  dist & 8 & 5 & 10 & 5737 & 14845 & 2.15e9 \\
\hline
\rule[-1ex]{0pt}{3.5ex}  f51m & 8 & 8 & 8 & 426 & 992 & 2.15e9 \\
\hline
\rule[-1ex]{0pt}{3.5ex}  mlp4 & 8 & 8 & * & * & * & *\\
\hline
\rule[-1ex]{0pt}{3.5ex}  clip & 9 & 5 & 11 & 15396 & 42127 & 2.16e9 \\
\hline
\rule[-1ex]{0pt}{3.5ex}  addm4 & 9 & 8 & * & * & * & * \\
\hline
\rule[-1ex]{0pt}{3.5ex}  b11 & 8 & 31 & * & * & * & * \\
\hline
\rule[-1ex]{0pt}{3.5ex}  apex4 & 9 & 19 & * & * & * & * \\
\hline
\rule[-1ex]{0pt}{3.5ex}  ex5 & 8 & 63 & * & * & * & * \\
\hline
\end{tabular}
\end{center}
\end{table}

\end{document}